# Time Variations of the Superkamiokande Solar Neutrino Flux Data by Rayleigh Power Spectrum Analysis

Koushik Ghosh and Probhas Raychaudhuri

*Abstract:* **We have used the Rayleigh Power Spectrum Analysis of the solar neutrino flux data from 1) 5-day-long samples from Super-Kamiokande-I detector during the period from June, 1996 to July, 2001; 2) 10 -day-long samples from the same detector during the same period and (3) 45-day long from the same detector during the same period. According to our analysis (1) gives periodicities around 0.25, 23.33, 33.75 and 42.75 months; (2) exhibits periodicities around 0.5, 1.0, 28.17, 40.67 and 52.5 months and (3) shows periodicities around 16.5 and 28.5 months. We have found almost similar periods in the solar flares, sunspot data, solar proton data ( $\bar{\epsilon}$ >10 Mev).**

## I. INTRODUCTION

Solar neutrino flux detection is very important not only to understand the stellar evolution but also to understand the origin of the solar activity cycle. Recent solar neutrino flux observed by Super-Kamiokande [1] and SNO detectors [2] suggest that solar neutrino flux from $^8$B neutrino and $^3$He+p neutrino from Standard Solar Model (S.S.M.) [3] is at best compatible with S.S.M. calculation if we consider the neutrino oscillation of M.S.W. [4] or if the neutrino flux from the sun is a mixture of two kinds of neutrino i.e. $\nu_e$ and $\nu_\mu$ [5]. Standard Solar Model (S.S.M.) are known to yield the stellar structure to a very good degree of precision but the S.S.M. cannot explain the solar activity cycle, the reason being that this S.S.M. does not include temperature and magnetic variability of the solar core [6,7]. The temperature variability implied a variation of the energy source and from that source of energy magnetic field can be generated which also imply a magnetic variability [7]. The temperature variation is important for the time variation of the solar neutrino flux. So we need a perturbed solar model and it is outlined by Raychaudhuri since 1971[6, 7], which may satisfy all the requirements of solar activity cycle with S.S.M.. For the support of perturbed solar model we have demonstrated that solar neutrino flux data are fractal in nature [8]. The excess nuclear energy from the perturbed nature of the solar model transformed into magnetic energy, gravitational energy and thermal energy etc. below the tachocline. The variable nature of magnetic energy induces dynamic action for the generation of solar magnetic field.

Recently Yoo *et al.* (2003)[9] searched the periodic modulations of the solar neutrino flux data of Super-Kamiokande-I (S.K.-I) detector from 31 May 1996 to 15 July 2001, almost half of the solar activity cycle, yielding a total detector life time of 1496 days. The solar neutrino data from S.K., acquired for 1871 elapsed days from the beginning of data are divided into roughly 10-day-long samples as listed in table I of Yoo *et al.* [9]. It is observed that not all of the data are perfectly of 10 days. They used Lomb periodogram method for unevenly arranged sample data to search for possible periodicities in the S.K.-I solar neutrino flux data. They have found no statistical significance of the periodicities in the S.K.-I solar neutrino flux data. However, Caldwell and Sturrock [10] used almost the same method i.e. Lomb-Scargle method of analysis and they have found a very interesting period of 13.75 days in the solar neutrino flux data of S.K.-I apart from other periods. Thus there arises a controversy regarding the periodicities of the solar neutrino flux data. Raychaudhuri [11] analysed the solar neutrino flux data of S.K.-I 45-days-sample data and have found 5 and 10 months period in the data and the same periods are also found in the $^{37}$Cl, SAGE and GALLEX solar neutrino flux data. 5 months period is seen in many solar activities (e.g. solar flares, sunspot etc.) indicating a relation between solar internal activities and solar surface activities.

The purpose of the paper is to see whether the Super-Kamiokande-I solar neutrino flux data is variable in nature or not. The observation of a variable nature of solar neutrino would provide significance to our understanding of solar internal dynamics and probably to the requirement of the modification of the Standard Solar Model i.e. a perturbed solar model. In this paper we shall study the solar neutrino flux data from 5-days-sample data, 10-days sample data and 45-days-sample data during the period from 31 May 1996 to 15 July 2001. We shall first study whether the data samples given by S.K.-I collaborators are random in nature or not. If

Koushik Ghosh is with the Department of Mathematics, Dr. B.C. Roy Engineering College, Durgapur-713206, India as a Lecturer (e-mail: koushikg123@yahoo.co.uk).

Probhas Raychaudhuri is with the Department of Applied Mathematics, University of Calcutta, 92, A.P.C Road, Calcutta-700009, India as a Professor (e-mail: probhasprc@rediffmail.com).

they are random then there may not be the possibility of any distribution of the data or any periodicities in the data of S.K.-I. If the data are non-random in nature then there is a possibility of periodicity in the S.K.-I data.

## II. STATISTICAL TEST OF RANDOMNESS

It is observed that S.K.-I solar neutrino flux data from 31 May 1996 to 15 July 2001 for 5-days, 10-days and 45-days data samples have large statistical errors from 15% to almost 30%. It is very difficult to evaluate precisely the statistical analysis of all the data. Without filtering we first evaluate the randomness of the data [12] from where Caldwell and Sturrock [10] confirmed their periodicities and Yoo et al. [9] have not found the periodicities in the S.K.-I data. We use the run test for the evaluation of data without filtering.

We have found that 5-days, 10-days and 45-days data of S.K.-I are random in nature while 30-days data evaluated from the 45-days data are not random. It is expected that 30-days data may have a period and we have found 5-months period with 99% confidence level.

It is already mentioned that original data of S.K.-I have errors 15% to 30%. So, it is necessary to smooth the data by filtering. The simplest filtering is the moving average method. So, we use 3 point moving average of the 5-days, 10-days and 45-days data of S.K.-I.

After making the moving average we have seen that all the data are non-random in nature and so the moving average data must follow a distribution. Hence the obtained moving average data can be satisfactorily used for the analysis of periodicity following the methods of Rayleigh Power Spectrum [13].

## III. RAYLEIGH POWER SPECTRUM ANALYSIS

Suppose we want to determine whether n events with angular values of $\{\theta_1, \theta_2, \theta_3, \ldots, \theta_n\}$ are uniformly distributed in angle. We can represent each event as a unit vector
$\vec{u}_i = \cos\theta_i \cdot \hat{e}_x + \sin\theta_i \cdot \hat{e}_y$ where $\hat{e}_x$ and $\hat{e}_y$ are unit vectors parallel to the x-axis and the y-axis respectively. The vector sum of these unit vectors is given by [13]

$$\vec{U} = \sum_{i=1}^{n} \cos\theta_i \cdot \hat{e}_x + \sum_{i=1}^{n} \sin\theta_i \cdot \hat{e}_y \qquad (1)$$

The magnitude of this vector divided by the number of events [13]

$$R = (1/n)[(\sum_{i=1}^{n}\cos\theta_i)^2 + (\sum_{i=1}^{n}\sin\theta_i)^2]^{1/2} \qquad (2)$$

indicates the uniformity of the distribution. If the events are uniformly distributed R is very close to zero. If on the other hand the events are concentrated around a certain angle, $\vec{R}$ is close to unity. The direction angle of the vector U shows the angle around which the events are concentrated. Bai and Cliver [13] defined the quantity Z as

$$Z = nR^2 = (1/n)[(\sum_{i=1}^{n}\cos\theta_i)^2 + (\sum_{i=1}^{n}\sin\theta_i)^2] \qquad (3)$$

for randomly distributed events and the distribution of Z follows [14] $P(Z>K)=\exp(-K)$. They obtained the "Rayleigh Power Spectrum" $Z(v)$ by setting $\theta_i = 2\pi t_i/T = 2\pi v_i$, where $\{t_i\}$ is a set of event occurrence times and T is a variable period [15].

It is to be noted that Bai and Cliver [13] did not consider the observed data of occurrence. They just considered the set of time of occurrence. Here we have modified the idea of Bai and Cliver [13] where we have considered the observed data as well as the set of time of occurrence. Here we have modified each event as a vector of modulus $|x(t_i)|$ instead of a unit vector considered by Bai and Cliver [13] as
$\vec{u}_i = x(t_i)\cdot\cos\theta_i\cdot\hat{e}_x + x(t_i)\cdot\sin\theta_i\cdot\hat{e}_y$ and the vector sum of these vectors is given by

$$\vec{U} = \sum_{i=1}^{n} x(t_i)\cos\theta_i\cdot\hat{e}_x + \sum_{i=1}^{n} x(t_i)\sin\theta_i\cdot\hat{e}_y \qquad (4)$$

Again the magnitude of this vector divided by the number of events is given by

$$R = (1/n)[(\sum_{i=1}^{n}x(t_i)\cos\theta_i)^2 + (\sum_{i=1}^{n}x(t_i)\sin\theta_i)^2]^{1/2} \qquad (5)$$

Here the quantity Z is defined as

$$Z = nR^2 = (1/n)[(\sum_{i=1}^{n}x(t_i)\cos\theta_i)^2 + (\sum_{i=1}^{n}x(t_i)\sin\theta_i)^2] \qquad (6)$$

We finally tabulate the considered T's and corresponding Z's for ultimate analysis. The values of T, which give significant peaks for Z, are considered to be the probable periods.

## IV. RESULTS

| Data | Periods (in months) obtained by Rayleigh Power Spectrum Analysis |
|---|---|
| | |
| (1) 5-day-long samples from Super-Kamiokande-I detector during the period from June 1996 to July 2001. | 0.25, 23.33, 33.75, 42.75. |
| (2) 10-day-long samples from Super-Kamiokande-I detector during the period from June 1996 to July 2001. | 0.5, 1.0, 28.17, 40.67, 52.5. |
| (3) 45-day-long samples from Super-Kamiokande-I detector during the period from June 1996 to July 2001. | 16.5, 28.5. |

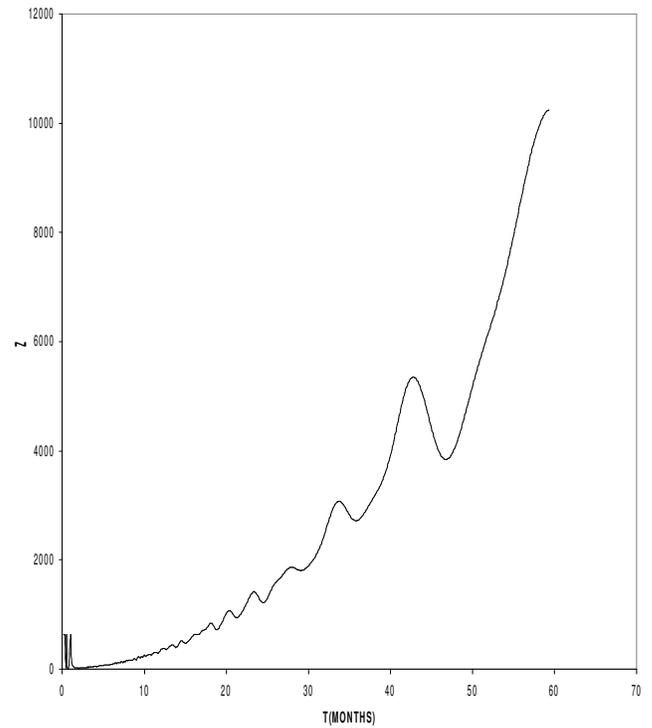

FIG.1: ANALYSIS OF 5-DAYS LONG SUPERKAMIOKANDE-I SOLAR NEUTRINO FLUX DATA BY RAYLEIGH POWER SPECTRUM ANALYSIS

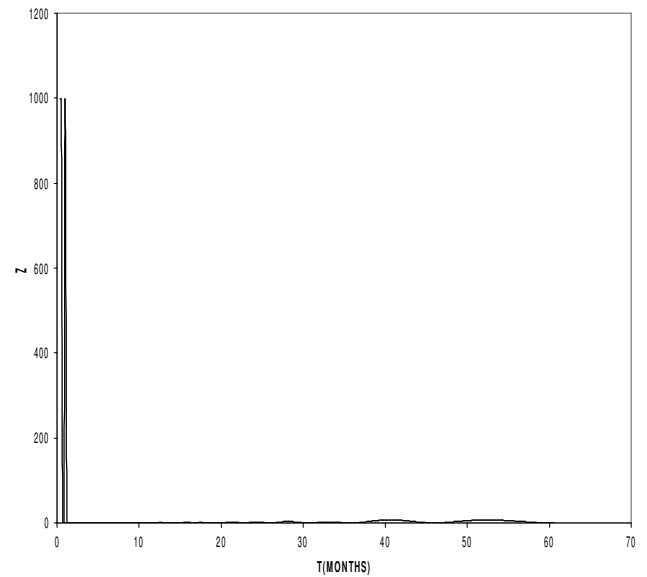

FIG.2: ANALYSIS OF 10-DAYS LONG SUPERKAMIOKANDE-I SOLAR NEUTRINO FLUX DATA BY RAYLEIGH POWER SPECTRUM ANALYSIS

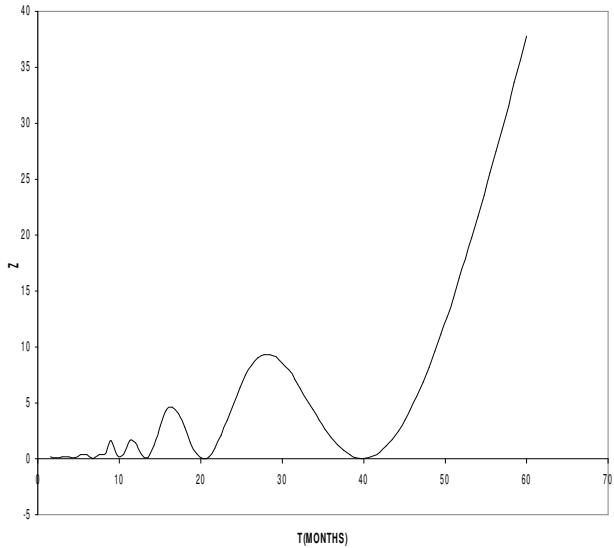

FIG.3: ANALYSIS OF 45-DAYS LONG SUPERKAMIOKANDE-I SOLAR NEUTRINO FLUX DATA BY RAYLEIGH POWER SPECTRUM ANALYSIS

## V. DISCUSSION

The observed period of 0.5 months in (2) is not appreciably different from the period of 13.75 days (at the frequency 26.57 $y^{-1}$) obtained by Caldwell and Sturrock [10] and the period of 13.76 days obtained by Yoo et al [9]. Moreover the obtained period of 0.25 months in (1) is significantly similar with the period of 0.21 months obtained for the same data by Ferraz-Mello method [16] and the period of 0.22 months obtained by Periodogram method [16]. The obtained period of 23.33 months in (1) is more or less similar with the period of 22.48 months obtained for the same data by Ferraz-Mello method [16] and the period of 24.02 months obtained by Periodogram method [16]. The obtained period of 33.75 months in (1) is almost similar with the period of 33.50 months obtained for the same data by Ferraz-Mello method [16]. The obtained period of 42.75 months in (1) falls very near to the period of 40.73 months obtained for the same data by Periodogram method [16]. The obtained period of 0.5 months in (2) is more or less similar with the period of 0.45 months obtained for the same data by Ferraz-Mello method [16] and the period of 0.39 months obtained by Periodogram method [16]. The obtained period of 1.0 month in (2) is almost similar with the period of 1.31 months obtained for the same data by Ferraz-Mello method [16]. The obtained period of 28.17 months in (2) is almost similar with the period of 32.99 months obtained for the same data by Ferraz-Mello method [16] and the period of 24.54 months obtained by Periodogram method [16]. The obtained period of 40.67 months in (2) is significantly similar with the period of 41.69 months obtained by Periodogram method [16]. The obtained period of 16.5 months in (3) is almost similar with the period of 14.01 months obtained for the same data by Ferraz-Mello method [16]. The obtained period of 28.5 months in (3) is more or less similar with the period of 32.50 months obtained for the same data by Ferraz-Mello method [16] and the period of 24.06 months obtained by Periodogram method [16].